\documentclass[aps,prd,twocolumn,superscriptaddress,showpacs,showkeys]{revtex4}
\usepackage{epsfig,epsf}
\usepackage{amsmath}
\usepackage{multirow}
\usepackage{booktabs}
\usepackage{simplewick}
\usepackage{slashed}
\usepackage{fancyhdr}
\usepackage{colonequals}
\usepackage{pstricks}
\usepackage[squaren]{SIunits}
\usepackage{color}
\usepackage{amssymb}
\usepackage{graphicx}
\usepackage{dsfont}
\usepackage{bbm}
\usepackage{textcomp}
\usepackage{amsfonts}
\usepackage{mathrsfs}
\usepackage[active]{srcltx}
\usepackage{remreset}
\makeatletter
\@removefromreset{figure}{chapter}
\renewcommand \thefigure{\@arabic\c@figure}
\makeatother

\newcommand{\DS}[1]{/\!\!\!#1}

\makeatletter
\def\fmslash{\@ifnextchar[{\fmsl@sh}{\fmsl@sh[0mu]}}
\def\fmsl@sh[#1]#2{%
  \mathchoice
    {\@fmsl@sh\displaystyle{#1}{#2}}%
    {\@fmsl@sh\textstyle{#1}{#2}}%
    {\@fmsl@sh\scriptstyle{#1}{#2}}%
    {\@fmsl@sh\scriptscriptstyle{#1}{#2}}}
\def\@fmsl@sh#1#2#3{\m@th\ooalign{$\hfil#1\mkern#2/\hfil$\crcr$#1#3$}}
\makeatother

\begin{document}

\title{Light-Cone Sum Rules for the $D_{(s)}\rightarrow \eta^{(\prime)} l \nu_l$ Form Factor}

\author{N.~Offen, F.A.~Porkert, A.~Sch\"afer}
\affiliation{Institut f\"ur Theoretische Physik, Universit\"at Regensburg,D-93040 Regensburg, Germany}

\begin{abstract}
We present an improved light-cone sum rule analysis of the decay form factors of $D$ and $D_s$ into 
$\eta$ and $\eta^{\prime}$ and argue that these decays offer a very promissing possibility to 
determine the leading Fock-state gluonic contribution of the $\eta'$ at future experimental facilities as FAIR or Super-KEKB. 
We also give the corresponding branching ratios for $B$ decays. 
\end{abstract}

\pacs{aaa}


\maketitle

\section{INTRODUCTION}
With the advent of high luminosity accelerators weak decays of hadrons 
containing valence charm or bottom quarks can be measured with 
very high precision. In fact, such decays might even offer one of
the best chances for 
the discovery of beyond the standard models physics, see the recent reviews  
\cite{Rosner:2012np,Stone:2012yr, Bediaga:2012py} and the citations given there. 
So, there is strong motivation to improve on the theoretical description
of the QCD input needed for such searches. 
One of the most important quantities for such exclusive channels are the hadron distribution 
amplitudes (DAs, often also called wave functions) and form factors. 
For each hadron DAs are characteristic nonperturbative quantities, just like PDFs. 
As for the latter, moments of DAs can be calculated on the lattice, see e.g. 
\cite{Arthur:2010xf,Braun:2008ur} and rapid progress can be expected along these lines.
Nevertheless 
input from many sides will be needed to understand in the long run the 
basic systematics of hadron DAs, even for the most important standard hadrons.
The controversial theoretical discussion spawned by the surprising {\sc BaBar} data for
the photon-pion transition form factor 
\cite{BABAR, Uehara:2012ag, Agaev:2010aq,Agaev:2012tm}
has illustrated that this field is still in a 
pioneering phase. 
Another non-perturbative approach, besides lattice QCD,  
to DAs and form factors are light-cone sum rules (LCSR) 
\cite{Balitsky:1989ry}. As both approaches are conceptually completely different 
the ideal situation is reached 
if both give the same results. We will show that this is what happens, e.g., 
for the decays $D_s\rightarrow \eta/\eta' + \ell + \nu_{\ell}$ we are analyzing
in this contribution. 
This case is especially interesting because the singlet-octet mixing of  
the $\eta$ and $\eta'$  should be reflected by the respective form factors, e.g. by a substantially 
different size of the gluonic contribution, see e.g. \cite{DiDonato:2011kr, Ricciardi:2012xu} for a recent review. 
As this debate is ongoing since many years it would be great news if the gluonic 
leading Fock-state contribution for the $\eta'$ could be experimentally determined.
(There always exist gluonic higher Fock-state components.)  
We will specify observables which are sensitive to this component and thus offer this oportunity.\\
From a theoretical point of view B-mesons 
would be better suited for our purpose. There the light-cone expansion exhibits a stronger hierarchy 
due to the larger mass of the b-quark which in turn reduces the uncertainty coming from the truncation of 
this expansion. However in practice this uncertainty is not the dominant one.\\
As for all three cases ($D$, $D_s$ and $B$ decays) 
the required increase in experimental accuracy looks very much feasible 
for next-generation experiments and we hope that in a few years from now  
data for this complete set of meson decays will provide undisputable  
experimental evidence for the gluonic component of the $\eta'$.\\
The decays  $D_s\rightarrow \eta/\eta' + \ell + \nu_\ell$ have been analysed 
before both phenomenologically, e.g.,  \cite{Ball:1995zv, Colangelo:2001cv} and using leading order LCSRs with chiral currents including meson 
mass corrections \cite{Azizi:2010zj}.
We improved that LO twist-2 analysis by taking into account all 
two-particle twist-2 and twist-3 NLO quark contributions and in addition the 
NLO twist-2 gluon contribution. The latter allows to extract information on the 
leading gluon DA of the $\eta'$. To achieve this goal we made heavy use of
NLO results existing in the literature
\cite{Ball:2007hb,Duplancic:2008ix}.
Our results for the decay formfactors agree within uncertainties with those of 
\cite{Azizi:2010zj}. While this is encouraging, we also feel that it is somewhat 
fortuitous, because we have some doubts concerning the benefits of the chiral currents 
used in that work, since they eliminate important nonperturbative information and do not couple only to the pseudoscalar mesons 
in the hadronic sum.\\
The decays $B\to\eta/\eta' + \ell + \nu_{\ell}$ were analysed in \cite{Aliev:2002tra} at leading order and in \cite{Ball:2007hb} at the same level of 
accuracy as in this note. We improve on the latter calculation by making an analysis of both the branching fractions and 
their ratios.

The paper is organized as follows: in section~II we discuss the $\eta - \eta^{\prime}$ mixing 
scheme used. In section~III  we outline the derivation of the LCSRs for the different 
form factors. In section~IV we present our numerical results and in
section~V we summarize and conclude.

\section{MIXING SCHEMES}
Two different schemes for describing the $\eta-\eta'$-mixing are commonly used: The singlet-octet (SO) \cite{Leutwyler:1997yr}
and the quark-flavour(QF)-scheme \cite{Feldmann:1998fks, Feldmann:1998sh, Feldmann:1998su, Feldmann:1999uf, Feldmann:2002kz}, see also \cite{DeFazio:2000my} 
for a mixing scheme independent sum rule determination of the couplings of the $\eta^{(\prime)}$ to the axial currents. 
The SO-scheme defines two hypothetical pure singlet and octet states 
$\vert \eta_{1,8}\rangle$ and two mixing angles $\Theta_{1,8}$ to describe the four decay constants 
\begin{equation}
 \left(
 \begin{array}{c c}   f_\eta^8&f_\eta^1\\f_{\eta^{\prime}}^8&f_{\eta^{\prime}}^1\end{array}\right)
~=~ 
\left(
\begin{array}{lr}
\cos\theta_8 & -\sin\theta_1 \\
\sin\theta_8 &  \cos\theta_1
\end{array}
\right)
\left(
\begin{array}{cc}
f_8 & 0 \\
0   & f_1
\end{array}
\right)
\end{equation}
defined as
\begin{equation}
 \langle 0\vert J_{\mu5}^i\vert P(p)\rangle=i f_P^i p_\mu,\quad (i=1,\,8,\;P=\eta,\,\eta').
\end{equation}
In this scheme $f_1$ describes the contribution of the $U(1)_A$-anomaly via the divergence of the singlet current $J_{\mu5}^1$ 
and the difference $\theta_i\neq0$ and $f_8\neq f_\pi$ is given by $SU(3)_F$-violating effects. $f_8$ and $\theta_i$ are scale independent 
and $f_1$ renormalises multiplicatively.\\
In the QF-scheme the basic currents and couplings are given by
\begin{eqnarray}
\langle 0 \rvert J_{\mu 5}^{a} \lvert \eta(p) \rangle &=:& i f_{\eta}^{a} p_{\mu},
\\
\langle 0 \rvert J_{\mu 5}^{a} \lvert \eta'(p) \rangle &=:& i f_{\eta'}^{a} p_{\mu},
\nonumber \\
J_{\mu 5}^{a}=&&
\left\{ \begin{array}{c r}
\frac{1}{\sqrt{2}}\left( \bar{u}\gamma_{\mu}\gamma_{5} u+\bar{d}\gamma_{\mu}\gamma_{5}d\right), & ~~~a=q \\
\bar{s}\gamma_{\mu}\gamma_{5} s,&  a=s.
\nonumber
\end{array}
\right.
\label{StromFqs}
\end{eqnarray}
Here the angles are scale dependent and their difference is given by OZI-rule violating contributions. Phenomenologically this difference 
is very small. 
Thus the authors of \cite{Feldmann:1998fks} proposed to use within the QF-scheme the approximation
\begin{equation}
 \phi\equiv\phi_{q,s},\quad \phi_q-\phi_s=0
\end{equation}
which has only three parameters with the phenomenological values 
\begin{eqnarray}
f_q&=&\left(1.07\pm0.02\right)f_{\pi}, 
\nonumber \\
f_s&=& \left(1.34\pm0.06\right)f_{\pi}, 
\nonumber\\ 
\phi&=&39.3\degree \pm 1.0 \degree,
\label{Eq:fs}
\end{eqnarray}
and where the mixing of the states follows the same pattern as for the decay constants:
\begin{equation}
\left( 
\begin{array}{l} 
\lvert \eta (p) \rangle \\ 
\lvert \eta^{\prime} (p) \rangle 
\end{array} 
\right)
~=~
\left( 
\begin{array}{l r} 
\cos\phi & -\sin\phi \\ 
\sin\phi & \cos\phi 
\end{array} 
\right)
\left( 
\begin{array}{c} 
\lvert \eta_q (p)\rangle \\ 
\lvert \eta_s (p)\rangle 
\end{array} 
\right)~~~.
\label{eq:state-mixing}
\end{equation}
The masses of the states to the order in which we perform our calculations are given by \cite{Feldmann:1998fks}
\begin{align}
m_{qq}^2&=m_\pi^2,\quad m_{ss}^2=2m_K^2-m_\pi^2~~~.\label{eq:masses} 
\end{align}
One important point to note is that in this version of the QF scheme there is no scale dependence left in the parameters. Since  
the mixing of the two different flavour states is given by OZI-rule violating contributions 
\begin{eqnarray}
\lvert \eta_q (p)\rangle &\propto& \phi_{2}^{q}(u)\lvert q\bar{q} \rangle + 
\phi_{2}^{\operatorname{OZI}}(u)\lvert s\bar{s} \rangle + \ldots,\\
\lvert \eta_s (p)\rangle &\propto& \phi_{2}^{\operatorname{OZI}}(u)\lvert q\bar{q} \rangle 
+ \phi_{2}^{s}(u)\lvert s\bar{s} \rangle + \ldots,
\end{eqnarray}
where
\begin{eqnarray}
 \phi_2^q&=&\dfrac{1}{3}(\phi_2^8+2\phi_2^1),\;\; \phi_2^s=\dfrac{1}{3}(2\phi_2^8+\phi_2^1),\nonumber\\
\phi_2^{\operatorname{OZI}}&=&\dfrac{\sqrt{2}}{3}(\phi_2^1-\phi_2^8)
\label{eq:da-osi-qf}
\end{eqnarray}
are leading twist distribution amplitudes, a consistent implementation requires to set 
\[\phi_2^{\operatorname{OZI}}=\dfrac{\sqrt{2}}{3}(\phi_2^1-\phi_2^8)=0.\]
This implies that one has to ignore the different scale-dependence of the singlet and octet distribution amplitudes, 
because otherwise their evolution would generate a non-zero $\phi_2^{\operatorname{OZI}}$. We followed \cite{Ball:2007hb} and set 
$\phi_2^1=\phi_2^8$ and evolved their lowest moment $a_2$ according to the octet scaling law. We confirm that the induced difference due to 
different renormalisation behavior is very small.
We also confirm their finding that the mixing of the leading Gegenbauer-moment in the conformal expansion of the twist 2 quark and gluon distribution amplitudes
\begin{eqnarray}
\phi_{2;\eta}^{1}(u,\mu)&=&6 u \bar{u} \left( 1 + \sum_{n=1}^{\infty} a_{2n}^{\eta,1}(\mu)C_{2n}^{3/2}(2u-1) \right),\nonumber \\
\psi_{2;\eta}^{g}(u,\mu)&=&u^2 \bar{u}^2 \sum_{n=1}^{\infty} B_{2n}^{\eta,g}(\mu)C_{2n-1}^{5/2}(2u-1) 
\label{GEGENBAUERexpansionGLUE}
\end{eqnarray}
given by \cite{Kroll:2002nt}
\begin{eqnarray}
\mu\frac{d}{d\mu} 
\left( 
\begin{array}{c}
a^{\eta,1}_2\\
B^{\eta,g}_2
\end{array}
\right)
&=&
\left( 
\begin{array}{cc}
\frac{100}{9} & -\frac{10}{81}\\
-36   &  22
\end{array}
\right)
\left( 
\begin{array}{c}
a^{\eta,1}_2\\
B^{\eta,g}_2
\end{array}
\right) ~~~.
\nonumber 
\end{eqnarray}
has only small numerical influence. This led us to neglect this effect in accordance with remarks made above.
Higher Gegenbauer-moments turned out to give only negligible contributions as well and therefore we restrict our analysis to the lowest moments. 
On the whole these effects are smaller than 3\%. The main difference with respect to \cite{Ball:2007hb}, besides using the $\overline{MS}$-mass for $m_c$, 
is that for the $D_s\to\eta^{(\prime)}$ decays we probe the $\bar{s}s$-content of the $\eta^{(\prime)}$ which leads to a different dependence on the mixing angle, see 
eq. (\ref{eq:qf-scheme-correlator}) while for the $D\to\eta^{(\prime)}$ the only difference is the change of Borel-parameter, 
continuum threshold and masses $m_c\leftrightarrow m_b$, $m_D\leftrightarrow m_B$.

\section{OUTLINE OF THE LCSR METHOD}

\begin{figure}[htbp]
\begin{center}
\includegraphics[width=.45\textwidth,clip=true]{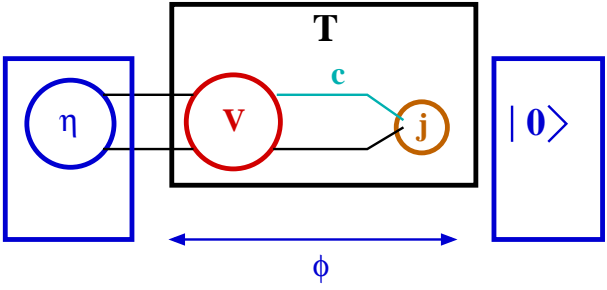}
\end{center}
\caption{Structure of the light cone sum rule calculation:
$j$ is the interpolating current for the heavy meson. 
The weak matrix element is contained in $V$ and the charm quark propagator is 
treated perturbatively. Thus, the factor $T$ can be calculated purely perturbatively, 
which is done at NLO accuracy. At that level the parton lines coupling into 
the $\eta$ can be either quark-antiquark or two gluons. The occurring matrix elements are 
parameterized in terms of the distribution amplitudes. A Borel transform serves to filter 
out the $D$ and $D_s$ contributions from $T$.  
}
\label{Fig_1}
\end{figure}

The idea behind LCSR calculations for decay matrix elements from heavy into light quark 
hadrons is illustrated in Fig.~\ref{Fig_1}.
For a detailed discussion of the original two-point sum rules and their extension consult, e.g., \cite{Shifman:1978by, Shifman:1978bx,
Shifman:1978bw,Belyaev:1994zk, Braun:1997kw, Colangelo:2000dp}.
In short one uses the two-fold nature of the correlation function to equate two different representations:
First one inserts a complete set of hadronic states, separates the ground state and expresses the rest 
via a dispersion integral over the hadronic spectral density.
Second one uses that for large negative virtualities the correlation function is dominated by light like distances and makes
an expansion around the light-cone leading to a convolution of perturbatively calculable hard scattering amplitudes and 
universal soft distribution amplitudes. After an analytic continuation of the light-cone expansion to physical momenta using 
a dispersion relation one equates these two representations by the assumption of quark-hadron duality. Finally it is customary 
to use a Borel transformation to suppress higher states in the hadronic sum and to get rid of subtraction terms which are necessary if the 
dispersion relation is divergent. We will illustrate these steps below.
Starting point for the $D_{\left(s\right)}^{+} \rightarrow \eta^{\left(\prime\right)} l^{+} \nu_{l}$ form factor 
\begin{eqnarray}
 \langle P(p)\rvert \bar{q}\gamma_\mu c\lvert D_{(s)}(p+q)\rangle&=&2 f_{D_{(s)}P}^+(q^2) p_\mu\nonumber\\
&+&\left(f_{D_{(s)}P}^+(q^2)+f_{D_{(s)}P}^-(q^2)\right)q_\mu\nonumber\\
\end{eqnarray}
is the correlation function:
\begin{align}
&F_{\mu}^{HP}\left(p,q\right)\nonumber\\
&=i\int d^{4}xe^{iq x}\,\langle P\left(p\right) \rvert \operatorname{T}\left\{ V_{\mu}^{P}\left(x\right),j^{\dagger}_{H}\hspace{-0.1mm}\left(0\right) \right\} \lvert 0 \rangle \nonumber\\
&=F^{HP}\hspace{-1mm}\left(q^2,\left(p+q\right)^2\right)p_{\mu}+\tilde{F}^{HP}\hspace{-1mm}\left(q^2,\left(p+q\right)^2\right)q_{\mu}.\nonumber\\
\label{CorrelationFunction}
\end{align}
where $P$ is the on-shell pseudoscalar meson, in our case
$P=\eta,\,\eta^{\prime}$, $H=B,\,D_{(s)}$, $V^P$ is the local weak interaction vertex 
and $j_H$ is a local interpolating current for the heavy quark system. 
In the present case we deal with the expressions collected in 
table~\ref{CurrentsTabular}. The scalar form factor 
\[f_{D_{(s)}P}^0(q^2)=f_{D_{(s)}P}^+(q^2)+\dfrac{q^2}{m_{D_{(s)}}^2-m_{\eta^{(\prime)}}^2}f_{D_{(s)}P}^-(q^2)\]
enters the leptonic spectrum only with factors proportional to $m_l^2$. Therefore we do not consider $\tilde{F}^P$ 
which is needed to calculate $f^0_{D_{(s)}\eta^{(\prime)}}$.

\vskip 0.3 cm
\begin{table}[htbp]
		\centering
		\begin{tabular}{ c  c  c }
		\toprule
	  \toprule
		Decay & interpolation current & weak current  \\
		\midrule
		$D_{s}^{+}\rightarrow \eta^{\left(\prime\right)} \, l \nu_{l}$ & $j_{D_{s}^{+}}=m_{c}\bar{s}i\gamma_{5}c$ & $V_{\mu}^{\left(\eta,\eta^{\prime}\right)}=\bar{s}\gamma_{\mu}c$ \\
		$D^{+}\rightarrow \eta^{\left(\prime\right)} \, l^{+} \nu_{l}$ & $j_{D^+}=m_{c}\bar{d}i\gamma_{5}c$ & $V_{\mu}^{\left(\eta,\eta^{\prime}\right)}=\bar{d}\gamma_{\mu}c$ \\
    \bottomrule
    \bottomrule
		\end{tabular}
		\caption{Currents entering the correlation function eq.~\eqref{CorrelationFunction}.}
		\label{CurrentsTabular}
\end{table}
Inserting a complete set of hadronic states between the two currents 
eq.~\eqref{CorrelationFunction}
and separating the ground state leads to

\begin{eqnarray}
 F^{D_{(s)}P}\hspace{-1mm}\left(q^2,\left(p+q\right)^2\right)&=&\dfrac{2 m_{D_{(s)}}^2f_{D_{(s)}} f_{D_{(s)}P}^+(q^2)}{(m_{D_{(s)}}^2-(p+q)^2)}\nonumber\\
&+&\int_{s_0^{h_{(s)}}}^\infty ds\dfrac{\rho^{h_{(s)}}(q^2,s)}{s-(p+q)^2},
\label{eq:had-side}
\end{eqnarray}
where $s_0^{h_{(s)}}$ is a hadronic threshold, $\rho^{h_{(s)}}(s)$ is the hadronic spectral density and $f_{D_{(s)}}$ is the decay constant of the $D(D_s)$-meson. 
Since the Borel transform will take care of subtraction terms in the end we won't write them anywhere.\\
The light-cone expansion for $q^2,(p+q)^2\ll m_c^2$ can be written in the general form
\begin{eqnarray}
 &&\left[F^{D_{(s)}P}(q^2,(p+q)^2))\right]_{OPE}=\sum_{t=2,3,4}F_0^{P,t}(q^2,(p+q)^2)\nonumber\\
&&+\dfrac{\alpha_s C_F}{4\pi}\sum_{t=2,3} F_1^{P,t}(q^2,(p+q)^2)+\cdots.
\end{eqnarray}
Here $t$ denotes the twist which is taken into account at the current accuracy. The leading and next to leading order 
expressions $F_{0,1}$ are given as convolutions of hard scattering amplitudes and distribution amplitudes, see 
figure \ref{Fig_1}:
\begin{align}
 &F_{0,1}^{D_{(s)}P,t}(q^2,(p+q)^2)
\nonumber\\
&=\int du ~T_{0,1}^{(t)}(q^2,(p+q)^2,m_c^2,u,\mu)~\phi_{\eta^{(\prime)}}^{(t)}(u,\mu).
\end{align}
$u$ denotes a generic expression for the momentum fractions of the partons in the meson and $\mu$ the factorisation scale. The leading order term is given 
by contracting the $c$-quarks to 
generate the free propagator and taking into account only the twist 2 distribution amplitude, see eq. (\ref{eq:2part}):
\begin{equation}
 F_0^{D_{(s)}\eta^{(\prime)},2}(q^2,(p+q)^2)=f_\eta m_c^2\int_0^1\dfrac{du\,\phi_{\eta^{(\prime)}}(u)}{m_c^2-q^2\bar{u}-(p+q)^2u}.
\end{equation}
Analytic continuation of the momentum $(p+q)^2$ flowing through the interpolating current leads to
\begin{eqnarray}
 &&\left[F^{D_{(s)}P}(q^2,(p+q)^2)\right]_{OPE}=\nonumber\\
&&\dfrac{1}{\pi}\int_{m_c^2}^\infty \dfrac{ds}{s-(p+q)^2}\,\mbox{Im}\,\left[F^{D_{(s)}P}(q^2,s)\right]_{OPE}~~~.
\label{eq:tw2-OPE}
\end{eqnarray}
Now the two representations can be equated by using the semi local quark-hadron duality assumption that from a certain continuum threshold $s_0^{D_{(s)}}$ on, the integral over 
the hadronic spectral density and over the partonic result should be the same:
\[\int_{s_0^{D_{(s)}}}^\infty ds\dfrac{\mbox{Im}\,\left[F^{D_{(s)}P}(q^2,s)\right]_{OPE}}{s-(p+q)^2}=\int_{s_0^{h_{(s)}}}^\infty ds\dfrac{\rho^{h_{(s)}}(q^2,s)}{s-(p+q)^2}.\]
This assumption and the final Borel transformation 
\[B_{M^2}\,\dfrac{1}{s-(p+q)^2}\longrightarrow e^{-\frac{s}{M^2}},\]
lead to the sum rule
\begin{eqnarray}
 f_{D_{(s)}P}^+(q^2)&=&\dfrac{1}{2m_{D_{(s)}}^2 f_{D_{(s)}}}e^{\frac{m_{D_{(s)}}^2}{M^2}}\nonumber\\
&\times&\dfrac{1}{\pi}\int_{m_c^2}^{s_0^{D_{(s)}}}ds\,\mbox{Im}\,\left[F^{D_{(s)}P}(q^2,s)\right]_{OPE}e^{-\frac{s}{M^2}},\nonumber\\
\label{eq:sum-rule-fin1}
\end{eqnarray}
where $M^2$ is the Borel-parameter. It is 
important to note that every additional two units of twist are accompanied by another power of the denominator
\begin{equation}
D=m_c^2-q^2\bar{u}-(p+q)^2u
\label{eq:tw-exp-par}
\end{equation}
which shows that for the processes in question the momentum transfer $q^2$ is severely constrained in order to have 
a converging light-cone expansion. Another point worth mentioning is that odd twists $3,5,\ldots$ come from the 
mass term of the c-quark propagator and are formally subleading in $\frac{1}{m_c}$ compared to their even counterparts. However, due to 
chiral enhancement coming from the prefactor $\mu_\eta$ of the twist 3 distribution amplitudes they numerically exceed these. 
This would imply 
that the unknown twist 5 contributions 
might be
larger than the twist 4 ones, which we analyse, and
convergence cannot be taken for granted. To really assess the situation  
a dedicated study of these higher twist contributions would be needed which is a
formidable task, far exceeding  
the scope of 
this note. To have at least a 
rough guess
of the resulting uncertainty 
we follow \cite{Khodjamirian:2009ys} and assume that the ratio of the unknown twist 5 term
to the twist 3 term is the same as the ratio of the twist 4 term and twist 2 term. This gives an additional 
uncertainty varying from $4\%$ for $q^2=-2$ GeV to $2.5\%$ for $q^2=0$.\\
The inclusion of the gluonic part of the $\eta^{(\prime)}$ in the sum rules was already discussed in \cite{Ball:2007hb} and we do not repeat it here. 
It boils down to using relation (\ref{eq:state-mixing}) to calculate the correlation functions
\begin{eqnarray}
F^{D_s\eta}&=&-F^{D_s\eta_s}\,\sin\phi+F^{D_s\eta_q}\,\cos\phi,\nonumber\\
F^{D_s\eta'}&=&F^{D_s\eta_s}\,\cos\phi+F^{D_s\eta_q}\,\sin\phi,\nonumber\\
F^{D\eta}&=&F^{D\eta_q}\,\cos\phi-F^{D\eta_s}\,\sin\phi,\nonumber\\
F^{D\eta'}&=&F^{D\eta_q}\,\sin\phi+F^{D\eta_s}\,\cos\phi
\label{eq:qf-scheme-correlator}
\end{eqnarray}
and insert these into equation (\ref{eq:sum-rule-fin1}).\\
The second summand in each equation of (\ref{eq:qf-scheme-correlator}) gets only contributions at NLO from the gluonic part while the first summand is a combination of quark 
and gluonic contributions. The quark contribution we take from \cite{Duplancic:2008ix, Khodjamirian:2009ys} with 
the replacements 
$f_\pi\to f_{q(s)}$, $f_\pi\frac{m_\pi^2}{2 m_q}\to f_{q}\frac{m_\pi^2}{2m_q},\,f_\pi \frac{m_\pi^2}{2 m_q}\to f_s\frac{2m_K^2-m_\pi^2}{2m_s}$ 
which means that we take $SU(3)$-flavour-violation into account only via the decay constants. In \cite{Khodjamirian:2009ys, Duplancic:2008tk} it was shown, that for 
decays into kaons and pions this is indeed a good approximation. We checked that our results do not change significantly if we include meson and quark mass corrections. 
But keeping all $SU(3)$-violating effects would force us to not only keep all quark- and meson mass dependences in the correlation function but to also use 
\begin{eqnarray}
h_q&=&f_q(m_\eta^2\cos^2\phi+m_{\eta'}^2\sin^2\phi)\nonumber\\&-&\sqrt{2}f_s(m_{\eta'}^2-m_\eta^2)\sin\phi\cos\phi,\nonumber\\
h_s&=&f_s(m_{\eta'}^2\cos^2\phi+m_\eta^2\sin^2\phi)\nonumber\\&-&\dfrac{f_q}{\sqrt{2}}(m_{\eta'}^2-m_\eta^2)\sin\phi\cos\phi,
\end{eqnarray}
\cite{Beneke:2002jn} instead of $f_q m_\pi^2$ and $f_s(2m_K^2-m_\pi^2)$ respectively. These quantities are, due to cancellations, very weakly constrained 
which would lead to uncertainties at the level of $200\%$ if one assumes uncorrelated errors in the twist 3 part, see e.g. \cite{Ball:2007hb}. 
In the ratios these uncertainties cancel for the largest part but for the form factors and decay rates this seems to be a huge overestimation.

\section{Numerics}

\subsection{CHOICE OF INPUT}
We follow \cite{Duplancic:2008ix, Khodjamirian:2009ys} in using the $\overline{MS}$-scheme
and one universal scale throughout our calculation. 
The scale is set to be 
$\mu \approx \sqrt{m_{D_{(s)}}^2 - m_c^2}=1.4 (1.5)$~ GeV 
and all quantities are evolved to this scale
using one-loop running for the quark masses 
and distribution amplitude parameters and two loop running 
for $\alpha_s$.\\
The values for the Gegenbauer-moments need some discussion. 
In a recent perturbative analysis \cite{Kroll:2012hs, Kroll:2013iwa} of the $\eta^{(\prime)}$ transition form factors 
P. Kroll and K. Passek-Kumeri\u{c}ki got the values (for $\mu=1\,\giga\electronvolt$)
\begin{equation}
 a_2^8=-0.05\pm 0.02,\quad a_2^1=-0.12\pm0.01,\quad a_2^g=19\pm5,
\label{eq:krollkum1}
\end{equation}
similar to their older results in \cite{Kroll:2002nt}, see also \cite{Ali:2003kg, Ali:2003ve}.
 Unfortunately, these numbers are at first sight in contradiction with 
the sum rule value  
$a_2^8\approx 0.2$. The authors 
of \cite{Kroll:2012hs, Kroll:2013iwa} state that 
there values are effective ones, 
contaminated by higher Gegenbauer-moments, while the effect of power corrections 
is neglected. Both effects were shown to be large in 
the accessible $Q^2$-region for the pion transition form factor in 
\cite{Agaev:2010aq, Agaev:2012tm}, where the value $a_2^\pi=0.13-0.16$
was obtained, in stark contrast  
to the value $a_2^\pi=-0.02\pm0.02$ obtained in \cite{Kroll:2012hs, Kroll:2013iwa}. 
Including generic power corrections lead to $a_2^8=0.06\pm0.05$ which also suggests that 
the values given in (\ref{eq:krollkum1}) should be taken with a grain of salt.
As we do not see how to correct for these effects we decided to ignore Eq. (\ref{eq:krollkum1})
and to use 
the average over sum rule fits to experimental data and direct lattice and sum rule 
calculations instead, leading to
\begin{equation}
 a_2^8(1\,\giga\electronvolt)=0.25\pm0.15.
\end{equation}
We implement the quark-flavour scheme by setting $a_2^1(1\,\giga\electronvolt)=a_2^8(1\,\giga\electronvolt)$ and evolving both via the renormalisation of 
the octet moment. This in turn implies $a_2^q=a_2^s$, see (\ref{eq:da-osi-qf}). As there is no hint for large $SU(3)$-flavour violation in the even Gegenbauer-moments, 
(one finds, e.g., $a_2^\pi\approx a_2^K$) which should be an acceptable approximation. Since the impact of the mixing between $a_2^1$ and $B_2^g$ is rather small, we treat the latter 
as a free parameter and vary it over the same very conservative range $B_2^g=0\pm20$ as in \cite{Ball:2007hb}. 
We take the quark- and meson masses from the Particle Data Group \cite{Beringer:1900zz}. Their current values are
\begin{align}
\overline{m}_c \left( \overline{m}_c \right)
&= \left( 1.275 \pm 0.025 \right) \giga\electronvolt,\\
m_{u}\left( \mu = 2 \giga\electronvolt \right) 
& = \left( 2.3^{+0.7}_{-0.5}\right) \mega\electronvolt, \\ 
m_{d}\left( \mu = 2 \giga\electronvolt \right) 
& = \left( 4.8^{+0.7}_{-0.3} \right) \mega\electronvolt, \\
m_{s}\left( \mu = 2 \giga\electronvolt \right) 
& = \left( 95 \pm 5 \right) \mega\electronvolt\,.
\end{align}
and
\begin{align}
m_{D^+}
&= 1869.6\mega\electronvolt,\qquad 
\!m_{D^+_s}
=& 1968.5\mega\electronvolt,\\ 
m_{\pi^{0}}
&= 134.98 \mega\electronvolt,\qquad 
m_{K^{0}}
=& 497.61\mega\electronvolt. 
\end{align}
The latter ones are related via flavour symmetry to the masses of the $\vert \eta_{q(s)}\rangle$-states as given in eq.~\eqref{eq:masses}.
For the pion decay constant we use $f_{\pi}= 130.4 \mega\electronvolt$, 
for the $D_{(s)}$ decay constant we take the experimental values from \cite{Beringer:1900zz}
\begin{align}
f_D \,\,   
&=\left(206.7 \pm8.5\pm2.5\right)\mega\electronvolt,\nonumber\\
f_{D_s}
&=\left(260   \pm5.4  \right)   \mega\electronvolt,\label{eq:Ddecayconstants}
\end{align}
while for the B-meson, in view of the existing large discrepancies in determinations 
of $\vert V_{ub}\vert$, which is in turn needed for 
the extraction of $f_B$, we use a two-point sum rule at order $\alpha_s$ \cite{Jamin:2001fw}. 
For the continuum threshold and the Borel-parameter we choose
\begin{align} 
s_{0}^{D}
&= \left(7\pm0.6\right)\giga\electronvolt^2,\nonumber\\
s_0^B
&= \left(35.75\pm 0.25\right) \giga\electronvolt^2,
\nonumber \\
M_{D_{(s)}}^2
&= \left(4.4\pm1.1\right)\giga\electronvolt^2,\nonumber\\
M_B^2
&= \left(18\pm3\right) \giga\electronvolt^2
\end{align}
and for the two point sum rule
\begin{align}
 \overline{s}_0^B
 &=\left(35.75\pm0.25\right)\giga\electronvolt^2,\nonumber\\
 \overline{M}_{\!B}^2
 &=\left(5\pm1\right)\giga\electronvolt^2
\end{align}
which fulfill the usual criteria for these parameters and are close to the ones used in \cite{Duplancic:2008ix, Khodjamirian:2009ys}.
The quark, gluon and mixed condensates are given by \cite{Ioffe:2005ym, Dosch:1997wb}
\begin{align}
\langle \bar{q}q \rangle \left( 2 \giga\electronvolt\right)
&= \left(-0.246_{+0.028}^{-0.019}\right)^3\giga\electronvolt^3,
\nonumber \\
\langle \tfrac{\alpha_S}{\pi} GG \rangle\left( 2 \giga\electronvolt\right)
&=
\left(0.012_{+ 0.006}^{-0.012}\right)\giga\electronvolt^4,\nonumber\\
m_0^2&=\dfrac{g\langle \bar{q} \sigma_{\mu\nu}G^{\mu\nu} q\rangle}{\langle\bar{q} q\rangle},\nonumber\\
&=\left(0.8\pm0.2\right)\giga\electronvolt^2 ~~~.
\end{align}
Finally we take for the the twist three and four parameters at $\mu=1 \giga\electronvolt$ 
\begin{eqnarray}
f_3^{\pi}&=&\left(0.0045\pm0.0015\right)\giga\electronvolt\squared,
\nonumber \\
\omega_3^{\pi}&=&\left(-1.5\pm0.7\right)\giga\electronvolt\squared,\nonumber\\
\epsilon_{\pi}&=&\left(\frac{21}{8}\right)\left(0.2\pm0.1\right)\giga\electronvolt\squared,
\nonumber \\
\delta_{\pi}^2&=&\left(0.18\pm0.06\right)\giga\electronvolt\squared.
\end{eqnarray}

\subsection{FORM FACTORS AND THEIR SHAPE}\label{ch:FFandShape}
As can be seen from eq. (\ref{eq:tw-exp-par}) our sum rules 
for $D$ and $D_s$ decays are only applicable for $q^2\ll m_c^2$. 
To be able to make a prediction for the shape of the form factor and for the value of the branching 
fractions we follow \cite{Khodjamirian:2009ys}: We calculate the form factors at $q^2<0$, where the 
twist expansion of the sum rules works perfectly well and then basically use a fit to extrapolate 
our results to $q^2>0$.
We use the simple Ball-Zwicky parametrization \cite{Ball:2004ye} having in mind that all fit formulas work nearly equally well \cite{Khodjamirian:2009ys, DescotesGenon:2008hh, Ball:2006wn}
and that unitarity constraints for more elaborate formulas are up to now not restrictive:
\begin{align}
f_{+}^{\operatorname{BZ}}(q^2) =& f_{+}(0)\left( \frac{1}{1-q^2/m_{D_{(s)}^{*}}^2} \right. \nonumber \\ 
& + \left. \frac{rq^2/m_{D_{(s)}^{*}}^2}{\left(1-q^2/m_{D_{(s)}^{*}}^2\right)\left(1-\alpha q^2/m_{D_{(s)}}^2\right)} \right)~~~.\label{eq:BZmodel}
\end{align}
The idea of this fit formula is basically to take the dispersive representation of the form factor, 
take out the known lowest lying resonance and approximate the dispersion integral over many particle 
states starting from $(m_{D_{(s)}}+m_\pi)^2$ by an effective pole. $r,\,\alpha$ parametrize the residuum and position of this 
pole, while $f_+(0)$ gives the overall normalisation. Despite the resonances $D_{(s)}^*$ being 
very close to the two-particle threshold the fits are numerically perfectly stable. 

\begin{figure}[htbp]
\includegraphics[width=.4\textwidth,clip=true]{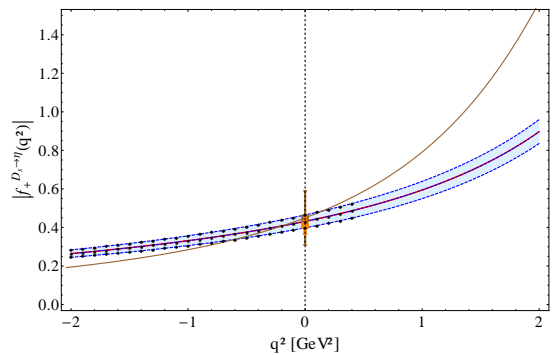}
\caption{$f_{+}^{D_s\eta'}(q^2)$ plotted as a function of $q^2$. The black dots are the calculated sum rule values. The blue straight line is the fit to the central values. 
Blue dashed band: Full uncertainties of our result. Red lines: Uncertainty coming from the gluonic contribution which due to a very small impact nearly conceal 
the blue line. Brown line: Results of \cite{Azizi:2010zj}. 
Orange Point: corrected lattice result from \cite{Kanamori:2013rha} in accordance with a private communication from the author.}
\label{DnachETAPlot}
\end{figure}

\begin{figure}[htbp]
\includegraphics[width=.4\textwidth,clip=true]{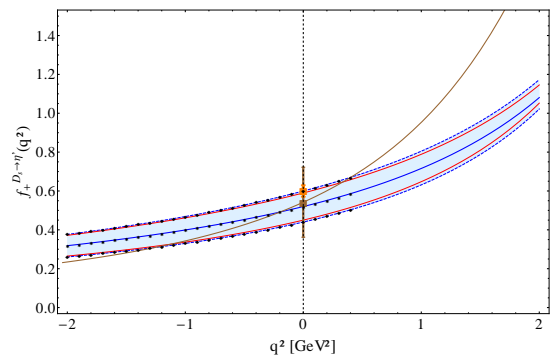}
\caption{$f_{+}^{D_s\eta^{\prime}}(q^2)$ plotted as a function of $q^2$. Same convention as in figure \ref{DnachETAPlot}}
\label{DnachETAprimePlot}
\end{figure}

The results for $f_+^{D_s\eta}(q^2)$ and $f_+^{D_s\eta'}(q^2)$ are shown in figure~\ref{DnachETAPlot} and figure~\ref{DnachETAprimePlot}. 
To get the error bands we made a statistical analysis of all input parameters at each $q^2\leq0$ assuming Gaussian uncertainties and then extrapolated them in the same way 
as the central values. As can be seen the uncertainty coming from the unknown gluon distribution amplitude is nearly negligible for 
the $f_+^{D_s\eta}(q^2)$ form factor which holds for $f_+^{D\eta}(q^2)$ and $f_+^{B\eta}(q^2)$ as well supporting the notion of a nearly 
total octet nature of the $\eta$. On the other hand there is a considerable impact on the $D_{(s)}(B)\to\eta'$-form factors from the gluonic part.
The fit-parameters can be found in Table \ref{tab:ShapeParametersEXP}. Figure \ref{DnachETAPlot} and \ref{DnachETAprimePlot} also contain 
results from a first lattice simulation for this quantity \cite{Kanamori:2013rha} which were corrected in accordance with a private communication from the author.
(The fact that one has to calculate disconnected contributions makes such lattice simulations very demanding \cite{Bali:2011yx})

\begin{table}[htbp]
		\centering
		\begin{tabular}{ c || c | c | c }
		\toprule
	  \toprule
		Decay &
		$r$ &
	  $\alpha$ & $\lvert f_{+}(0) \rvert$
	  \\
		\midrule
		$D_{s}^{+}\rightarrow \eta \, l^{+} \nu_{l}$ &
		$0.284^{+0.003}_{-0.002}$ &
		$0.252^{+0.107}_{-0.082}$ &
		$0.432^{+0.033}_{-0.033}$
		\\
		$D_{s}^{+}\rightarrow \eta^{\prime} l^{+} \nu_{l}$ &
		$0.284^{+0.137}_{-0.095}$ &
		$0.252^{+0.382}_{-0.395}$ &
		$0.520^{+0.080}_{-0.080}$
		\\
		$D^{+}\rightarrow \eta \, l^{+} \nu_{l}$ &
		$0.174^{+0.001}_{-0.001}$ &
		$-0.043^{+0.068}_{-0.052}$ &
		$0.552^{+0.051}_{-0.051}$
		\\
		$D^{+}\rightarrow \eta^{\prime} l^{+} \nu_{l}$ &
		$0.174^{+0.243}_{-0.142}$ &
		$-0.043^{+0.526}_{-0.596}$ &
		$0.458^{+0.105}_{-0.105}$ 
		\\
    \bottomrule
    \bottomrule
		\end{tabular}
		\caption{Shape parameters for $f_{+}^{D_{(s)}^{+}\hspace{-0.5mm}\eta^{\left(\prime\right)}}\hspace{-1.5mm}(q^2)$ as input for the BZ-model eq.~\eqref{eq:BZmodel} 
.}
		\label{tab:ShapeParametersEXP}
\end{table}

\begin{figure}[htbp]
\includegraphics[width=.4\textwidth,clip=true]{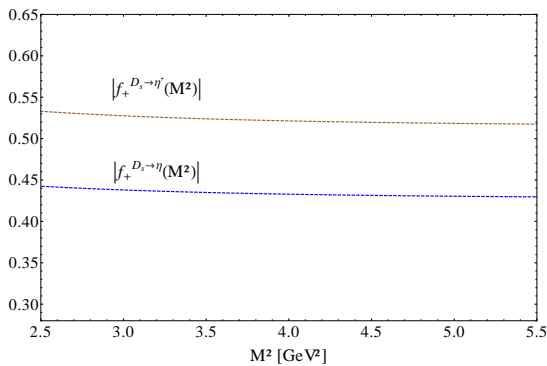}
\caption{
$\vert f_{+}^{D_s\eta}(q^2=0)\vert$, $\vert f_{+}^{D_s\eta^{\prime}}(q^2=0)\vert$ plotted as a function of the Borel parameter $M^2$. 
The blue dashed line corresponds to $\vert f_{+}^{D_s\eta}(q^2=0)\vert$ 
and the brown dashed line to $\vert f_{+}^{D_s\eta^{\prime}}(q^2=0)\vert$.}
\label{DnachETABorelPlot}
\end{figure}

Our results for $q^2=0$ are shown in Table \ref{tab:fplus0}. For illustration we show the dependence of the $D_s\to\eta^{(\prime)}$ form factors on the Borel-parameter in figure~\ref{DnachETABorelPlot}.

As can be seen the sum rules are stable for a very large range of parameter values. 
\squeezetable\begingroup
\begin{table*}
\caption{\it Form factors $f^{D_s^+ \eta^{\left(\prime\right)}}_+(0)$, $f^{D^+\eta^{\left(\prime\right)}}_+(0)$ and $f^{B^+\eta^{\left(\prime\right)}}_+(0)$
calculated from LCSRs eq.\eqref{eq:qf-scheme-correlator}.}
\begin{ruledtabular}
\begin{tabular}{cccccccccc}
\hline$\vphantom{\displaystyle\int}$
Formf. centr.value
&$M^2$
&$\mu$
&$\left(s_0^D/s_0^B\right)$
&$a_2$
&$B_2^g$
&$\left(fq,\, fs,\, \phi \right)$
&twist-3
&twist-4
&$\left(\text{condensates},\,m_c/m_b \right)$\\
\hline$\vphantom{\displaystyle\int}$
$\lvert f^{D^+_s \eta}_+(0) \rvert = 0.432$
&$\pm 0.003$
&$\pm 0.026$
&$\pm 0.010$
&$\pm 0.013$
&$\pm 0.001$
&$\pm 0.025$
&$\pm 0.014$
&$\pm 0.002$
&$\pm 0.005$
\\
\hline$\vphantom{\displaystyle\int}$
$\lvert f^{D^+_s \eta^{\prime}}_+ \hspace{-0.75mm} (0)\rvert = 0.520$ 
&$\pm 0.003$
&$\pm 0.032$
&$\pm 0.012$
&$\pm 0.015$
&$\pm 0.070$
&$\pm 0.028$
&$\pm 0.016$
&$\pm 0.002$
&$\pm 0.006$
\\
\hline$\vphantom{\displaystyle\int}$
$\lvert f^{D^+ \eta}_+(0) \rvert = 0.552$
&$\pm 0.008$
&$\pm 0.034$
&$\pm 0.013$
&$\pm 0.016$
&$\pm 0.002$
&$\pm 0.015$
&$\pm 0.036$
&$\pm 0.002$
&$\pm 0.007$
\\
\hline$\vphantom{\displaystyle\int}$
$\lvert f^{D^+ \eta^{\prime}}_+ \hspace{-0.75mm} (0)\rvert = 0.458$
&$\pm 0.007$
&$\pm 0.028$
&$\pm 0.011$
&$\pm 0.013$
&$\pm 0.096$
&$\pm 0.025$
&$\pm 0.030$
&$\pm 0.002$
&$\pm 0.006$
\\
\hline$\vphantom{\displaystyle\int}$
$\lvert f^{B^+ \eta}_+ (0)\rvert = 0.238$
&$\pm 0.002$
&$\pm 0.013$
&$\pm 0.002$
&$\pm 0.004$
&$\pm 0.001$
&$\pm 0.006$
&$\pm 0.011$
&$\pm 0.0002$
&$\pm 0.007$
\\
\hline$\vphantom{\displaystyle\int}$
$\lvert f^{B^+ \eta^{\prime}}_+ \hspace{-0.75mm} (0)\rvert = 0.198$
&$\pm 0.001$
&$\pm 0.011$
&$\pm 0.002$
&$\pm 0.003$
&$\pm 0.061$
&$\pm 0.007$
&$\pm 0.009$
&$\pm 0.0001$
&$\pm 0.006$
\\
\hline
\end{tabular}
\end{ruledtabular}
\label{tab:fplus0}
\end{table*}
\endgroup

\begin{figure}[htbp]
\begin{center}
\includegraphics[width=.4\textwidth,clip=true]{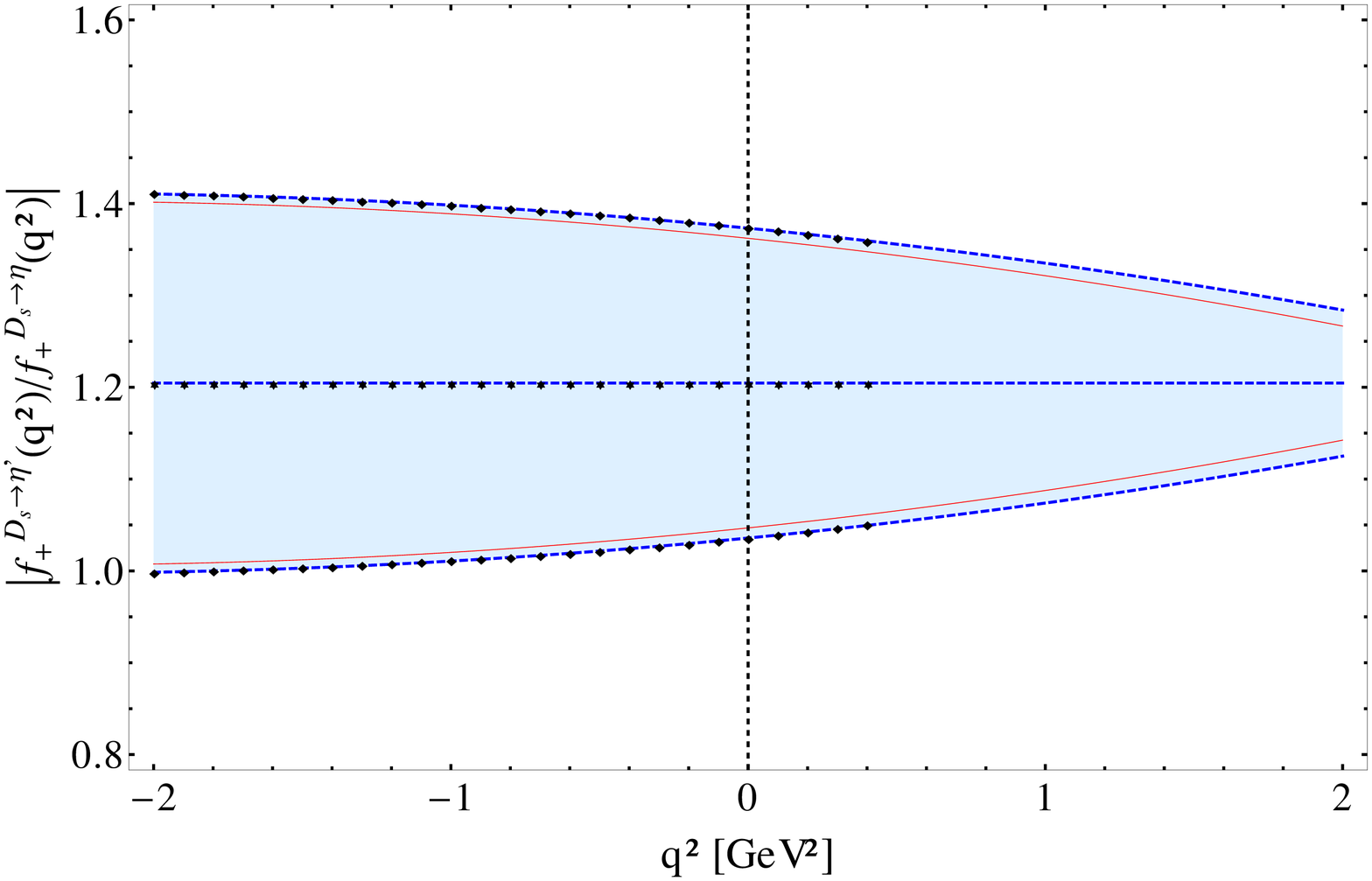}
\end{center}
\caption{
$\lvert f_{+}^{D_s\eta^{\prime}}(q^2)/f_{+}^{D_s\eta}(q^2) \rvert$ plotted as a function of $q^2$. Again the black dots are the calculated sum rule values. 
The blue straight line is the fit to the central values, while 
the blue dashed band corresponds to the uncertainties of our result. It is completely dominated by 
the gluonic contribution (red dashed lines).
}
\label{fig:GEGENBAUERgluePlot}
\end{figure}

Especially interesting are the ratios of the $\eta'$ to $\eta$ form factors since for such ratios most of the uncertainties cancel. For the gluonic part we made the assumption 
$B_2^{g,\eta}=B_2^{g,\eta'}$ since no large $SU(3)$-breaking is expected in this Gegenbauer-moment. Note however that the contribution to the form factors 
is vastly different, due to the different admixture of the singlett part which is given by the dacay constants

\begin{eqnarray}
  f_\eta^1&=&\sqrt{\dfrac{2}{3}}\cos \phi\,f_q-\sqrt{\dfrac{1}{3}}\sin \phi\,f_s,\nonumber\\
 f_{\eta'}^1&=&\sqrt{\dfrac{2}{3}}\sin \phi\,f_q+\sqrt{\dfrac{1}{3}}\cos \phi\,f_s,
\end{eqnarray}
see eq. (\ref{eq:2part-glue}), (\ref{eq:f_1-qf}).\\
What can be seen from table (\ref{tab:fplusDIV}) is that almost the whole uncertainty comes 
from $B_2^g$ which would give the possibility to constrain this quantity if more precise experimental data would be available. 
The result for the $D_s$-form factors in the considered $q^2$-region is shown in figure \ref{fig:GEGENBAUERgluePlot}. 
As can be seen the uncertainties are completely governed by the gluonic contribution. Table \ref{tab:fplusDIV} shows our results at $q^2=0$.

\squeezetable\begingroup
\begin{table*}
\caption{\it Ratios $\left|\tfrac{f^{D_{(s)}^+ \eta^{\prime}}_+(0)}{f^{D_{(s)}^+\eta}_+(0)}\right|$ and $\left|\tfrac{f^{B^+ \eta^{\prime}}_+(0)}{f^{B^+\eta}_+(0)}\right|$
calculated from LCSRs eq.\eqref{eq:qf-scheme-correlator}.}
\begin{ruledtabular}
\begin{tabular}{c c c c c c c c c c}\hline$\vphantom{\displaystyle\int}$
Formf. centr.value
&$M^2$
&$\mu$
&$\left(s_0^D/s_0^B\right)$
&$a_2$
&$B_2^g$
&$\left(fq,\, fs,\, \phi \right)$
&twist-3
&twist-4
&$\left(\text{condensates},\,m_c/m_b \right)$\\
\hline$\vphantom{\displaystyle\int}$
$\left|\tfrac{f^{D_{s}^+ \eta^{\prime}}_+(0)}{f^{D_{s}^+\eta}_+(0)}\right| = 1.20$
&$\pm 1\cdot10^{-13}$
&$\pm 1\cdot10^{-12}$
&$\pm 6\cdot10^{-13}$
&$\pm 7\cdot10^{-14}$
&$\pm 0.16$
&$\pm 0.06$
&$\pm 3\cdot10^{-12}$
&$\pm 3\cdot10^{-14}$
&$\pm 2\cdot10^{-14}$
\\
\hline$\vphantom{\displaystyle\int}$
$\left|\tfrac{f^{D^+ \eta^{\prime}}_+(0)}{f^{D^+\eta}_+(0)}\right| = 0.83$
&$\pm 5\cdot10^{-13}$
&$\pm 9\cdot10^{-13}$
&$\pm 2\cdot10^{-13}$
&$\pm 5\cdot10^{-15}$
&$\pm 0.18$
&$\pm 0.04$
&$\pm 8\cdot10^{-13}$
&$\pm 3\cdot10^{-14}$
&$\pm 5\cdot10^{-14}$
\\
\hline$\vphantom{\displaystyle\int}$
$\left|\tfrac{f^{B^+ \eta^{\prime}}_+(0)}{f^{B^+\eta}_+(0)}\right| = 0.83$ 
&$\pm 8\cdot10^{-13}$
&$\pm 6\cdot10^{-13}$
&$\pm 1\cdot10^{-13}$
&$\pm 1\cdot10^{-13}$
&$\pm 0.26$
&$\pm 0.04$
&$\pm 8\cdot10^{-13}$
&$\pm 2\cdot10^{-14}$
&$\pm 2\cdot10^{-13}$
\\
\hline
\end{tabular}
\end{ruledtabular}
\label{tab:fplusDIV}
\end{table*}
\endgroup

\subsection{Branching fractions and experimental results}
With an extrapolation of the form factors to the whole kinematic region we are able to calculate the 
branching fractions and compare them to experimental results. For massless leptons the scalar form factor $f^0_{D_{(s)}\eta^{(\prime)}}$ 
does not contribute so the decay rate is given by
\begin{align}
&\Gamma\left( D_{s}^+ \rightarrow \eta^{\left(\prime\right)} l^+ \nu_{l} \right) \nonumber \\
& = \frac{G_F^2\lvert V_{cs} \rvert^2}{24\pi^3}\int_{0}^{q_{\operatorname{max}}^2}\mathrm dq^2\mathrm\,\lambda^{\frac{3}{2}}(q^2)\lvert f_{+}^{D_s^{+}\hspace{-0.5mm}\eta^{\left(\prime\right)}}\hspace{-1.5mm}(q^2) \rvert^2, \label{eq:Grartio}
\end{align}
where the kinematical function $\lambda(q^2)$ is defined via 
\begin{align}
\lambda(x)&=\frac{1}{4 m_{H}^{2}}\left[\left(m_{H}^{2}+m_{M}^{2}-x\right)^2-4m_{H}^{2}m_{M}^{2}\right],
\end{align}
with ($P=\eta,\,\eta^{\prime}$; $H=B^{+},D^{+},D_s$). After multiplication with the mean life time of the 
considered meson we get the relevant branching fractions. To extract the uncertainties we again assume Gaussian errors 
and extrapolate the error of $\vert f_+(q^2)\vert^2$ 
with different fit functions from $q^2<0$ to the physical region. The deviations found due to the 
change of the fit function are incorporated in the error budget. Our results and 
the experimental values are shown in table \ref{tab:BranchingFractions}.

\begin{table}[ht]
\begin{center} 
\begin{tabular}{l||l|l c}
\hline  \hline
Decay & LCSRs (this work)&Experiment\\\hline\hline
   $D_s \to \eta^{\prime} e \nu_e$
&  $(0.75\pm 0.23)\%$ 
&  $(0.91\pm 0.33)\%$
&\cite{Yelton:2009aa}\\\hline
   $D_s \to \eta e \nu_e$
&  $(2.00 \pm 0.32)\%$ 
&  $(2.48 \pm 0.29)\%$
&\cite{Yelton:2009aa}\\\hline
   $D \to \eta^{\prime} e \nu_e$
&  $(3.86 \pm 1.77)\cdot 10^{-4}$ 
&  $\left( 2.16 \pm 0.53\pm 0.07 \right)\cdot 10^{-4}$ &\cite{Yelton:2010js} \\\hline
   $D \to \eta e \nu_e$ 
&  $(24.5 \pm 5.26)\cdot 10^{-4}$ 
&  $\left( 11.4\pm 0.9\pm 0.4 \right)\cdot 10^{-4}$
&\cite{Yelton:2010js} \\\hline
   $B \to \eta^{\prime} e \nu_e$
& 
&  $(2.66 \pm 0.80 \pm 0.56) \cdot 10^{-4}$
&\cite{Adam:2007pv}\\
& \raisebox{1.75ex}[-1.75ex]{ $(0.36 \pm 0.22)\cdot 10^{-4}$} 
&  $(0.24 \pm 0.08 \pm 0.03) \cdot 10^{-4}$
&\cite{delAmoSanchez:2010zd}\\\hline
   $B \to \eta e \nu_e$
&
&  $(0.44 \pm 0.23 \pm 0.11) \cdot 10^{-4}$
&\cite{Adam:2007pv}\\
&\raisebox{1.75ex}[-1.75ex]{ $(0.73 \pm 0.20)\cdot 10^{-4}$} 
&  $(0.36 \pm 0.05 \pm 0.04) \cdot 10^{-4}$
&\cite{delAmoSanchez:2010zd}\\\hline
 \end{tabular}
\caption{Branching fractions for the different decays.\label{tab:BranchingFractions}} 
\end{center}
\end{table}
Again the ratios turn out to be especially interesting since most of the uncertainties in the theoretical calculation cancel and 
they are dominated by the contribution of the gluonic Gegenbauer-moment $B_2^g$. Here we made the same assumption $B_2^{g,\eta}=B_2^{g,\eta'}$ 
as for the ratios of the form factors. Comparing them to the experimental values,
\begin{eqnarray}
\frac{\Gamma\left(D^+_s \to \eta^{\prime} e^+ \nu_e\right)}{\Gamma\left(D^+_s \to \eta e^+ \nu_e\right)} 
&=& 0.37 \pm 0.09 \,(B_2^g) \pm 0.04 \,(\text{rest}),\nonumber\\
&Exp:&0.36 \pm 0.14\quad\mbox{\cite{Yelton:2009aa}},
\nonumber\\
\frac{\Gamma\left(D^+ \to \eta^{\prime} e^+ \nu_e\right)}{\Gamma\left(D^+ \to \eta e^+ \nu_e\right)}
&=& 0.16 \pm 0.06\,(B_2^g) \pm 0.02\,(\text{rest}),\nonumber\\
&Exp:&0.19 \pm 0.09\quad\mbox{\cite{Yelton:2010js}},
\nonumber\\
\frac{\Gamma\left(B \to \eta^{\prime} e^+ \nu_e\right)}{\Gamma\left(B \to \eta e^+ \nu_e\right)}
&=& 0.50 \pm 0.29\,(B_2^g) \pm 0.05\,(\text{rest}),\nonumber\\
&Exp:&0.67 \pm 0.24 \pm 0.1\quad\mbox{\cite{delAmoSanchez:2010zd}},
\label{Eq:42}
\end{eqnarray}
one can see good overall agreement
but, as can clearly be seen, the experimental precision is up to now not sufficient to draw any conclusion on 
$B_2^g$.


\section{SUMMARY AND DISCUSSION}

We have calculated the form factors and branching fractions of the decays $D_{(s)}\to\eta^{(\prime)} l\nu$ and $B\to\eta^{(\prime)} l\nu$ in the framework of 
light cone sum rules for massless leptons. The form factors were shown to agree with 
available lattice results and the branching ratios 
eq. (\ref{Eq:42}) 
with experiment.
So the overall picture is nicely consistent. 
Our main result is, however, the error budget given in eq. (\ref{Eq:42})
clearly showing that $B_2^g$ dominates the uncertainties in all cases. Therefore, 
already a moderate increase in experimental accuracy will allow to determine the gluonic contribution 
to $\eta$ and $\eta'$ from all three ratios, providing a sensitive consistency check.
FAIR and Super-KEKB should even provide precision measurements of these ratios and thus 
allow to settle this long-standing issue.

\section*{Acknowledgements}
We thank I. Kanamori for providing us an update of the results of \cite{Kanamori:2013rha}. 
This work was supported by Forschungszentrum J\"ulich
(FFE contract 42008319).
\begin{widetext}
\section*{Appendix}
\section*{Definitions of distribution amplitudes}
Here we give the definitions of the used distribution amplitudes. We follow the notation of \cite{Ball:2006wn}, see also 
\cite{Khodjamirian:2009ys} for a minor correction, for the quark-antiquark
\begin{eqnarray}
 &&\langle \eta(p)|\bar{q}_\omega^i(x_1) 
q^j_\xi(x_2)|0\rangle_{x^2\to 0}
  = \frac{i\delta^{ij}}{12}f_\eta 
\int_0^1 du~e^{iu p\cdot x_1 +i\bar{u}p\cdot x_2}
\Bigg ( [\DS p \gamma_5]_{\xi\omega} \varphi_\eta(u)
\nonumber\\
&&\qquad\qquad -[\gamma_5]_{\xi\omega}\mu_\eta\phi^p_{3\eta}(u)
+\frac 16[\sigma_{\beta\tau}\gamma_5]_{\xi\omega}p_\beta(x_1-x_2)_\tau \mu_\eta\phi^\sigma_{3\eta}(u)
\nonumber\\
&&\qquad\qquad  +\frac1{16}[\DS p \gamma_5]_{\xi\omega}(x_1-x_2)^2\phi_{4\eta}(u)
-\frac{i}{2} [(\DS x_1-\DS x_2) \gamma_5]_{\xi\omega}
\int\limits_0^u\psi_{4\eta}(v)dv 
\Bigg)
\label{eq:2part}
\end{eqnarray}
and quark-antiquark-gluon distributions
\begin{eqnarray}
&& \langle \eta(p)|\bar{q}_\omega^i(x_1) g_sG_{\mu\nu}^a(x_3)
q^j_\xi(x_2)|0
\rangle_{x^2\to 0}
=\frac{\lambda^a_{ji}}{32}\int {\cal D}\alpha_ie^{ip(\alpha_1 x_1+\alpha_2 x_2+\alpha_3x_3)}
\nonumber
\\
&& \times\Bigg[if_{3\eta}(\sigma_{\lambda\rho}\gamma_5)_{\xi
\omega}(p_\mu p_\lambda g_{\nu\rho}-
p_\nu p_\lambda g_{\mu\rho})\Phi_{3\eta}(\alpha_i)
\nonumber
\\
&& -f_\eta(\gamma_\lambda\gamma_5)_{\xi\omega}\Big\{(p_\nu 
g_{\mu\lambda}-p_\mu g_{\nu\lambda})\Psi_{4\eta}(\alpha_i)
+\frac{p_\lambda(p_\mu x_\nu-p_\nu x_\mu)}{(p\cdot x)}
\left(\Phi_{4\eta}(\alpha_i)+\Psi_{4\eta}(\alpha_i)\right)
\Big\}
\nonumber
\\
&& -\frac{if_\eta}2\epsilon_{\mu\nu\delta\rho}(\gamma_\lambda)_{\xi\omega}
\Big\{(p^\rho g^{\delta\lambda}-p^\delta g^{\rho\lambda})
\widetilde{\Psi}_{4\eta}(\alpha_i)+  
\frac{p_\lambda(p^\delta x^\rho-p^\rho x^\delta)}{(p\cdot x)}
\left(\widetilde{\Phi}_{4\eta}(\alpha_i)+
\widetilde{\Psi}_{4\eta}(\alpha_i)\right)\Big\}\Bigg]\,.
\nonumber 
\\
\label{eq:3part}
\end{eqnarray}
For the gluon-gluon distribution amplitude we take over the notation of \cite{Ball:2007hb}
\begin{eqnarray}
\langle \eta^{(\prime)}(p) | G_{\mu x}(x)[x,-x] \tilde G^{\mu x}(-x) | 0 \rangle &=& 
f_{\eta^{(\prime)}}^1\frac{C_F}{2\sqrt{3}} (px)^2 ~ \int_0^1 du e^{-i \left(2u-1\right) px} \psi_{2,\eta^{(\prime)}}^g(u)~~~.
\label{eq:2part-glue}
\end{eqnarray}
with
\begin{eqnarray}
 f_\eta^1&=&\sqrt{\dfrac{2}{3}}\cos \phi\,f_q-\sqrt{\dfrac{1}{3}}\sin \phi\,f_s,\nonumber\\
 f_{\eta'}^1&=&\sqrt{\dfrac{2}{3}}\sin \phi\,f_q+\sqrt{\dfrac{1}{3}}\cos \phi\,f_s
\label{eq:f_1-qf}
\end{eqnarray}
which differs by a normalisation factor of $\sigma=\sqrt{\frac{3}{C_F}}$ to the one used in \cite{Kroll:2002nt}.
The explicit conformal expansion of the different distribution amplitudes can be found in \cite{Ball:2006wn, Khodjamirian:2009ys, Kroll:2002nt, Ball:2007hb}
and we don't write them here in order not to lengthen this note further.
\end{widetext}

\end{document}